\documentclass[preprintnumbers,article,amsmath,amssymb,floatfix,10pt,prd,onecolumn,
superscriptaddress,nofootinbib]{revtex4-2}
\usepackage{bm}
\usepackage{amsfonts}
\usepackage{latexsym}
\usepackage[latin1]{inputenc}
\usepackage{graphicx}
\usepackage{amsmath}
\usepackage{palatino}
\usepackage{mathpazo}
\usepackage{textcomp}
\usepackage{float}
\usepackage{booktabs}
\usepackage{dcolumn}
\usepackage{ragged2e}
\usepackage{hyperref}
\hypersetup{colorlinks,citecolor=blue}
\hypersetup{colorlinks=true,linkcolor=blue,filecolor=magenta,    urlcolor=blue}
\usepackage{amsmath}
\usepackage{subfigure}
\usepackage[]{natbib}
\usepackage{xcolor}
\usepackage{orcidlink}
\usepackage{epsfig}
\usepackage{caption}
\usepackage{subcaption}
\usepackage{commath}
\def\jnl@style{\it}
\def\aaref@jnl#1{{\jnl@style#1}}

\def\aaref@jnl#1{{\jnl@style#1}}

\def\aj{\aaref@jnl{AJ}}                   
\def\apj{\aaref@jnl{ApJ}}                 
\def\apjl{\aaref@jnl{ApJ}}                
\def\apjs{\aaref@jnl{ApJS}}               
\def\apss{\aaref@jnl{Ap\&SS}}             
\def\aap{\aaref@jnl{A\&A}}                
\def\aapr{\aaref@jnl{A\&A~Rev.}}          
\def\aaps{\aaref@jnl{A\&AS}}              
\def\mnras{\aaref@jnl{Mon.~Not.~Roy.~Astron.~Soc.}}             
\def\prd{\aaref@jnl{Phys.~Rev.~D}}        
\def\prc{\aaref@jnl{Phys.~Rev.~C}}  
\def\prl{\aaref@jnl{Phys.~Rev.~Lett.}}    
\def\qjras{\aaref@jnl{QJRAS}}             
\def\skytel{\aaref@jnl{S\&T}}             
\def\ssr{\aaref@jnl{Space~Sci.~Rev.}}     
\def\zap{\aaref@jnl{ZAp}}                 
\def\nat{\aaref@jnl{Nature}}              
\def\aplett{\aaref@jnl{Astrophys.~Lett.}} 
\def\apspr{\aaref@jnl{Astrophys.~Space~Phys.~Res.}} 
\def\physrep{\aaref@jnl{Phys.~Rep.}}      
\def\physscr{\aaref@jnl{Phys.~Scr}}       
\def\commat{\aaref@jnl{Comm.~Math.~Phys.}}              
\def\science{\aaref@jnl{Science}}               
\def\cqg{\aaref@jnl{Classical Quant.~Grav.}}            
\def\jpcs{\aaref@jnl{JPCS}}                                     
\def\ijmpd{\aaref@jnl{Int.~J.~Mod.~Phys.~D}}                    
\def\grg{\aaref@jnl{Gen.~Relat.~Gravit.}}               
\def\rpp{\aaref@jnl{Rep.~Prog.~Phys.}}          
\def\npa{\aaref@jnl{Nucl.~Phys.~A}}        
\def\lrr{\aaref@jnl{Living Rev.~Rel.}}                   
\def\jcap{\aaref@jnl{J.~Cosmology Astropart.~Phys.}}    
\def\rmp{\aaref@jnl{Rev.~Mod.~Phys.}}   
\def\epjc{\aaref@jnl{Eur.~Phys.~J.~C}} 
\def\plb{\aaref@jnl{~Phy.~Lett.~B}} 
\def\mpla{\aaref@jnl{Mod.~Phy.~Lett.~A}} 
\def\arxiv{\aaref@jnl{arxiv.org}}

\allowdisplaybreaks[1]

\begin{document}
\color{black}       
\title{Observational constraints on the equation of state of viscous fluid in $f(R, T)$ gravity}

\author{M. Koussour\orcidlink{0000-0002-4188-0572}}
\email[Email: ]{pr.mouhssine@gmail.com \textcolor{black}{ (Corresponding author)}}
\affiliation{Department of Physics, University of Hassan II Casablanca, Morocco.}

\author{A. Altaibayeva}
\email[Email: ]{aziza.ltaibayeva@gmail.com \textcolor{black}{ (Corresponding author)}}
\affiliation{Department of General and Theoretical Physics, L.N. Gumilyov Eurasian National University, Astana 010008, Kazakhstan.}

\author{S. Bekov}
\email[Email: ]{ss.bekov@gmail.com}
\affiliation{Department of General and Theoretical Physics, L.N. Gumilyov Eurasian National University, Astana 010008, Kazakhstan.}
\affiliation{Kozybayev University, Petropavlovsk, 150000, Kazakhstan.}

\author{O. Donmez\orcidlink{0000-0001-9017-2452}}
\email[Email: ]{orhan.donmez@aum.edu.kw}
\affiliation{College of Engineering and Technology, American University of the Middle East, Egaila 54200, Kuwait.}

\author{S. Muminov}
\email[Email: ]{sokhibjan.m@urdu.uz}
\affiliation{Urgench State University, Kh. Alimjan str. 14, Urgench 221100, Uzbekistan.}

\author{J. Rayimbaev\orcidlink{0000-0001-9293-1838}}
\email[Email: ]{javlon@astrin.uz}
\affiliation{University of Tashkent for Applied Sciences, Gavhar Str. 1, Tashkent 100149, Uzbekistan.}
\affiliation{National University of Uzbekistan, Tashkent 100174, Uzbekistan.}


\begin{abstract}
In this paper, we investigate a cosmological model based on viscous $f(R,T)$ gravity as a potential alternative to dark energy. This model incorporates bulk viscosity and is analyzed using an effective equation of state. We consider the simplest specific model, $f(R,T)=R+\lambda T$, where $\lambda$ is a constant. The exact solution of our viscous $f(R,T)$ cosmological model is derived, and then we use the combined datasets consisting of 31 $H(z)$ data points and 1701 Pantheon+ SNe data points to determine the best-fit values of the model parameters. We find good agreement with observations, particularly at higher redshifts. Our model's behavior, including energy density, pressure with viscosity, effective equation of state, and deceleration parameter, is analyzed. It indicates a shift from decelerated to accelerated phases of the universe's expansion, suggesting that bulk viscosity in the cosmic fluid could effectively generate the negative pressure necessary for cosmic expansion. Finally, we explore statefinder diagnostics to differentiate between various dark energy models, revealing that our model resides in the quintessence region.\\

\end{abstract}

\maketitle

\tableofcontents

\section{Introduction}
\label{sec1}

Cosmology underwent a significant transformation with the confirmation of the accelerating expansion of the universe through observational evidence from Type Ia supernova (SNe Ia) searches \cite{Riess,Perlmutter}. Over the past two decades, numerous observational findings, including those from large-scale structure (LSS) \cite{E11}, the Wilkinson Microwave Anisotropy Probe (WMAP) \cite{D.N.}, the cosmic microwave background radiation (CMBR) \cite{C.L.,R.R.}, and baryonic acoustic oscillations (BAOs) \cite{D.J.,W.J.}, have consistently supported the notion of cosmic acceleration. The leading explanation for this acceleration is the existence of dark energy (DE), which is described by an equation of state (EoS) $\omega_0= -1.018 \pm 0.057$ in a flat cosmos \cite{Planck/2020}. Another promising approach to explain the accelerating expansion of the universe without invoking undetected DE is to consider a more general gravitational action. These cosmological models modify the Einstein-Hilbert (EH) action of general relativity (GR) by introducing a generic function $f(R)$, where $R$ denotes the Ricci scalar curvature. This concept was first proposed in \cite{H.A.,R.K.,H.K.}. The $f(R)$ gravity can explain the expansion of the universe without the need for exotic DE components \cite{Carr,Cap}. Observational signatures of $f(R)$ gravity, including constraints from the solar system and the equivalence principle, are discussed in \cite{Shin,Sal,Alex}. Viable cosmological models of $f(R)$ gravity that pass solar system tests have been proposed \cite{Noj,V.F.,L.A.}. Odintsov et al. \cite{Odi-1,Odi-2} investigated the Hubble constant tension and the influence of energy conditions in models of $f(R)$ gravity. Refs. \cite{Noj-2,Noj-3,Noj-4,Odi-3,Odi-4} provide detailed insights into the diverse implications of $f(R)$ gravity cosmological models.

An extension of curvature-based $f(R)$ gravity, involving an explicit coupling of the generic function $f(R)$ with the matter Lagrangian density $L_m$, was introduced in \cite{O.B.}. Harko and Lobo introduced a theory of $f(R, L_m)$ gravity \cite{THK-6}, which extends matter-curvature coupling theories by considering a generic function $f(R, L_m)$ dependent on the Ricci scalar $R$ and the matter Lagrangian $L_m$. In 2011, Harko et al. \cite{THK-7} introduced $f(R, T)$ gravity through a general non-minimal coupling between matter and geometry. In this modified gravity, the EH action is replaced by a functional $f(R, T)$ instead of just $R$, where $T = g^{\mu \nu} T_{\mu \nu}$ represents the trace of the energy-momentum tensor. This theory has garnered significant research interest recently \cite{Myrzakulov/2012,Houndjo/2012,Barrientos/2014,Vacaru/2014}. Jamil et al. \cite{Jamil/2012} reconstructed several cosmological models within this theory of gravity using the functional form $f(R, T)=R^2+g(T)$. Sharif and Zubair \cite{Sharif/2012,Sharif/2014} studied a perfect fluid distribution and a massless scalar field within the context of the Bianchi type-I universe. Recently, da Silva et al. \cite{Silva} investigated the properties of rapidly rotating neutron stars using $f(R,T)$ gravity. They studied the impact of $f(R,T) = R + 2\lambda T$ gravity on the structure and characteristics of neutron stars, which are known for their extreme density and compactness. Vinutha et al. \cite{Vinutha} analyzed the field equations and derived the dynamical equations for anisotropic perfect fluid cosmological models in $f(R,T)$ gravity. They studied the solutions and examined the effects of anisotropy on the evolution of the scale factor, energy density, and other cosmological parameters. Bishi et al. \cite{Bishi} explored the existence of the G\"{o}del Universe in various functional forms of $f(R,T)$ gravity. Their investigation aimed to find cosmological solutions resembling the G\"{o}del universe, which features rotation and closed timelike curves.

In this study, we investigate $f(R,T)$ cosmology coupled with a viscous fluid. Viscosity in cosmology refers to the study of how viscosity, a measure of a fluid's resistance to deformation, affects the behavior of cosmic fluids on large scales. When a cosmic fluid undergoes rapid expansion, it deviates from thermodynamic equilibrium, leading to the generation of an effective pressure. The high viscosity observed in cosmic fluids can be understood as a manifestation of this effective pressure \cite{J.R.,H.O.}. There are two key viscosity coefficients in cosmology: shear viscosity $\eta$ and bulk viscosity $\zeta$. Shear viscosity is associated with velocity gradients in the fluid. When assuming a universe described by a homogeneous and isotropic Friedmann-Lem\^aitre-Robertson-Walker (FLRW) metric, shear viscosity is typically neglected. However, when the FLRW metric assumption is relaxed, several cosmological models incorporating shear viscosity fluid have been developed. These models are discussed in various studies, such as \cite{bali/1988,bali/1987,deng/1991,huang/1990,Momeni/2012}. In contrast, bulk viscosity, which is the focus here, introduces damping due to volumetric straining. The concept involves incorporating the bulk viscosity coefficient $\zeta$ into the $f(R,T)$ gravity. We assume that $\zeta$ follows a scaling law and transforms the Einstein case into a proportional form for the Hubble parameter. This scaling law has been demonstrated to be highly beneficial. Basically, modified gravity theories provide the rationale for cosmic expansion, while viscosity coefficients play a crucial role in the pressure term, leading to cosmic acceleration. Samanta et al. \cite{samanta/2017} investigated Kaluza-Klein bulk viscous fluid cosmological models within the framework of $f(R, T)$ gravity. They examined the validity of the second law of thermodynamics in these models, shedding light on the interplay between gravity, viscosity, and thermodynamic principles in cosmology. Satish and Venkateswarlu \cite{satish/2016} explored cosmological models of bulk viscous fluids in the context of the anisotropic Kaluza-Klein universe within the framework of $f(R, T)$ gravity. Recently, Sadatian \cite{Davood/2019} investigated the effects of viscous content on the modified cosmological $f(T)$ model. The study likely explores how viscosity influences the dynamics of the universe in the context of modified teleparallel gravity theories. Moreover, Srivastava and Singh \cite{Srivastava/2018} proposed a new holographic DE model within the framework of modified $f(R,T)$ gravity theory. The model incorporates constant bulk viscosity, offering a novel approach to understanding DE phenomena. To explore additional cosmological models incorporating bulk viscous fluid, one can refer to works such as \cite{beesham/1993,colistete/2007,Brevik/2012,Singh/2014,Fabris/2006,Brevik/2017,Mohan/2017,Debnath/2019,Gadbail/2021,Mahichi/2023}. Furthermore, interesting applications of bulk viscous cosmology in the context of black holes are discussed in \cite{bb1,bb2}.

The paper is structured as follows: In Sec. \ref{sec2}, we introduce the action and fundamental formulation that govern the dynamics within $f(R,T)$ gravity. Sec. \ref{sec3} outlines the equations of motion within the flat FLRW universe. In Sec. \ref{sec4}, we consider a functional form $f(R, T)$ and derive the cosmological solutions in terms of the Hubble parameter. 
In Sec. \ref{sec5}, we conduct an analysis of observational data to establish the best-fit values of the model parameters using datasets including the Hubble $H(z)$ datasets (31 points) and the Pantheon+ datasets (1701 samples). Further, in Sec. \ref{sec6}, we examine the behavior of various cosmological parameters, including energy density, effective pressure, effective EoS parameter, and deceleration parameter. In Sec. \ref{sec7}, we investigate the behavior of statefinder parameters, aiming to distinguish between different DE models. Finally, our results are discussed in Sec. \ref{sec8}.

\section{Overview of $f(R,T)$ theory}
\label{sec2}

The $f(R, T)$ theory represents a modified approach to gravity, where the most comprehensive action for $f(R, T)$ gravity is expressed as \cite{THK-7}
\begin{equation}
\mathbb{S}=\frac{1}{2}\int  f(R,T)\sqrt{-g}d^{4}x +\int \mathcal{L}_{m}\sqrt{-g}d^{4}x.\label{action}
\end{equation}

Here, we set $8\pi G = 1$ and $c = 1$. The function $f(R, T)$ is arbitrary, with $R$ representing the Ricci scalar, $T$ indicating the trace of the energy-momentum tensor $T=g_{\mu\nu}T^{\mu\nu}$, and $L_m$ denotes the Lagrangian density of the matter field. The energy-momentum tensor of matter is expressed as 
\begin{equation}\label{1e}
T_{\mu\nu} = \frac{-2}{\sqrt{-g}} \frac{\delta(\sqrt{-g}L_m)}{\delta g^{\mu\nu}}.
\end{equation}

Furthermore, the Ricci scalar $R$ can be derived by contracting the Ricci tensor $R_{\mu\nu}$ as
\begin{equation}\label{1b}
R= g^{\mu\nu} R_{\mu\nu}.
\end{equation} 

The Ricci tensor is defined as
\begin{equation}\label{1c}
R_{\mu\nu}= \partial_\lambda \Gamma^\lambda_{\mu\nu} - \partial_\mu \Gamma^\lambda_{\lambda\nu} + \Gamma^\lambda_{\mu\nu} \Gamma^\sigma_{\sigma\lambda} - \Gamma^\lambda_{\nu\sigma} \Gamma^\sigma_{\mu\lambda},
\end{equation}
where $\Gamma^\alpha_{\beta\gamma}$ denotes the components of the Levi-Civita connection.

By varying the action (\ref{action}) with respect to the metric tensor $g_{\mu\nu}$, we derive the following field equation that governs the dynamics of gravitational interactions,
\begin{equation}
\label{field}
f_{R}(R,T)R_{\mu\nu}-\frac{1}{2}f(R,T)g_{\mu\nu}+(g_{\mu\nu}\Box -\nabla _{\mu}\nabla
_{\nu})f_{R}(R,T) = T_{\mu\nu}-f_{T}(R,T)T_{\mu\nu}- f_{T}(R,T)\Theta _{\mu\nu},
\end{equation}
where $f_R(R,T) \equiv \frac{\partial f(R,T)}{\partial R}$, $f_{T}(R,T) \equiv \frac{\partial f(R,T)}{\partial T}$, and $\Theta_{\mu\nu}$ is defined as
\begin{equation}
\Theta_{\mu\nu} \equiv g^{\alpha\beta}\frac{\delta T_{\alpha\beta}}{\delta g^{\mu\nu}}=-2T_{\mu\nu} +g_{\mu\nu}L_m -2g^{\alpha\beta}\frac{\partial^2 L_m}{\partial g^{\mu\nu}\partial g^{\alpha \beta}}.
\end{equation}

From Eq. (\ref{field}), it is clear that the behavior of the field equations in $f(R,T)$ gravity is determined by the physical properties of the matter field. Thus, selecting different matter sources will result in various cosmological models within $f(R,T)$ gravity. Put differently, by choosing different forms of the functional $f(R,T)$, one can create viable cosmological models. Harko et al. \cite{THK-7} have proposed three models by defining the functional form $f(R,T)$ as either $f(R,T) = R + 2f(T)$, $f(R,T) = f_1(R) + f_2(T)$, or $f(R,T) = f_3(R) + f_4(R) f_5(T)$, where $f(T)$, $f_1(R)$, $f_2(T)$, $f_3(R)$, $f_4(R)$, and $f_5(T)$ are arbitrary functions of $R$ and $T$. These functions can be selected arbitrarily, and the resulting outcomes can be compared with observations related to late-time acceleration or reconstructed based on plausible physical principles, such as cosmic thermodynamics and energy conditions. In this paper, we examine the function $f(R, T) = R + 2f(T)$, where $f(T)$ represents an arbitrary function of the trace of the energy-momentum tensor of matter. The inclusion of $f(T)$ modifies the gravitational interaction between matter and curvature. Specifically, for the choice $f(R, T) = R + 2f(T)$, the gravitational field equation gives
\begin{equation}
 R_{\mu\nu}-\frac{1}{2}R g_{\mu\nu} =T_{\mu\nu}+f(T) g_{\mu\nu}-2 f_T \left[T_{\mu\nu} + \Theta_{\mu\nu} \right],
 \label{field}
\end{equation}
where $f_T$ represents the derivative of $f(T)$ with respect to $T$. 

Based on the cosmological principle, the universe is homogeneous and isotropic on large scales, implying that the distribution of matter is uniform and there are no preferred directions in space \cite{ryden/2003}. In the next section, we will explore the implications of this principle by applying the gravitational field equations of $f(R,T)$ gravity to a homogeneous and isotropic universe, further elucidating the dynamics of cosmic evolution within the framework of this modified theory of gravity.

\section{Equations of motion in $f(R, T)$ gravity}
\label{sec3}

To investigate the cosmological implications, we adopt the homogeneous and spatially isotropic FLRW metric \cite{ryden/2003},
\begin{equation}
ds^2=dt^2-a^2 (t) \left[dr^2+ r^2 (d\theta^2+sin^2\theta d\phi^2 ) \right], \label{FLRW}
\end{equation} 
where $a(t)$ represents the scale factor of the universe. Also, the Ricci scalar derived for metric (\ref{FLRW}) is
\begin{equation}\label{2b}
R= -6 ( \dot{H}+2H^2 ),
\end{equation}
where $H=\frac{\dot{a}}{a}$ denotes the Hubble parameter, representing the rate of expansion of the universe.

Now, we will consider a bulk viscous fluid, and here we provide some reasons supporting this choice. Firstly, incorporating bulk viscosity in a fluid can be viewed as an effort to refine its description, reducing its idealized properties. This is evident in realistic models of stellar astrophysics, as detailed in \cite{gusakov/2007,gusakov/2008,haensel/2002}. In conditions of spatial homogeneity and isotropy, which align with the cosmological principle, bulk viscous pressure emerges as the sole admissible dissipative phenomenon. In a gas dynamical model, the presence of an effective bulk pressure can be attributed to a non-standard self-interacting force acting on the gas particles \cite{colistete/2007}. The presence of bulk viscosity leads to a negative contribution to the total pressure, as evidenced in \cite{odintsov/2020,fabris/2006,meng/2009}. Because of spatial isotropy, the bulk viscous pressure is uniform in all spatial directions and thus proportional to the volume expansion rate, $\theta = 3H$. The effective pressure of the cosmic fluid is given by \cite{brevik/2005,gron/1990,C.E./1940}
\begin{equation}
    \bar{p}=p-\zeta \theta=p-3\zeta H,
\end{equation}
where $p$ represents the standard pressure, and $\zeta > 0$ denotes the bulk viscosity coefficient. Here, we consider $\zeta$ as a free parameter in our model.

The associated energy-momentum tensor is expressed as
\begin{equation}\label{2c}
T_{\mu\nu}=(\rho+\bar{p})u_\mu u_\nu - \bar{p}g_{\mu\nu},
\end{equation}
where $\rho$ represents the matter-energy density, $u^\mu$ is the four-velocity components, and $u^\mu u_\mu=1$. The matter Lagrangian density can be represented as $L_m =-\bar{p}$, and the trace of the energy-momentum tensor is expressed as $T=\rho-3\bar{p}$. Thus, the expression of  $\Theta_{\mu\nu}$ is given by 
\begin{equation}
 \Theta_{\mu\nu}= -2 T_{\mu\nu}- \bar{p}\;g_{\mu\nu}.
 \label{theta}
\end{equation}

The relationship between the standard pressure and the matter-energy density is given by $p=(\gamma-1)\rho$, as shown in \cite{Ren/2006}, where $\gamma$ is a constant within the range $0 \leq \gamma \leq 2$. Therefore, the effective equation of state (EoS) for the viscous fluid can be expressed as
\begin{equation}\label{2e}
\bar{p}= (\gamma-1)\rho -3\zeta H.
\end{equation}

Using Eqs. (\ref{field}) and (\ref{theta}), the gravitational field equations in $f(R, T)$ gravity can be expressed as 
  \begin{equation}
 R_{\mu\nu}-\frac{1}{2}R g_{\mu\nu} =T_{\mu\nu}+2f_T T_{\mu\nu}+\left[2 \bar{p} f_T+ f(T)\right] g_{\mu\nu}.
\end{equation}

In addition, it is important to note that the covariant divergence of the matter-energy-momentum tensor in the context of the $f(R, T)$ theory can be expressed as
\begin{equation}
    \nabla ^{\mu }T_{\mu \nu }=-\frac{1 }{1+ f_{T} }%
\left[ T_{\mu \nu }\nabla ^{\mu }f_{T} +g_{\mu \nu }\nabla
^{\mu }\left( f_{T} \bar{p}\right) \right]. 
\end{equation}

Therefore, the equation above demonstrates that in the $f(R, T)$ theory, the matter-energy-momentum tensor is not conserved. The lack of conservation in the matter energy-momentum tensor implies the existence of an additional force affecting massive test particles, leading to non-geodesic motion. Physically, it represents the energy flow into or out of a defined volume of a physical system. Furthermore, the presence of a non-zero right-hand side of the energy-momentum tensor suggests the occurrence of transfer processes or particle production within the system. Notably, the energy-momentum tensor is conserved when there are no $f_T$ terms in the equation \cite{THK-7}.

\section{Cosmological Solutions}
\label{sec4}

In our study of cosmological models involving bulk viscosity fluid, we focus on a simple model obtained by selecting the function $f(T)$ such that $f(T) = \lambda T$, where $\lambda$ is a constant serving as a coupling parameter between geometry and matter \cite{Sahu17,Mishra16}. Then, for this particular functional form, the modified Friedmann equations describing the universe dominated by bulk viscous matter in $f(R, T)$ gravity are given by
\begin{equation}
\label{F1}    
3H^2=(1+3\lambda)\rho-\lambda \bar{p},
\end{equation}
and
\begin{equation}
2\dot{H}+3H^2=\lambda \rho-(1+3\lambda)\bar{p}.
\label{F2}
\end{equation}

It is worth noting that the standard cosmological models with viscosity can be retrieved from modified Friedmann equations (\ref{F1})-(\ref{F2}) for $\lambda = 0$. Now, by using Eqs. (\ref{F1}) and (\ref{F2}), the matter-energy density can be expressed as,
\begin{equation}
\rho =\frac{(3+6\lambda)H^2-2\lambda \dot{H}}{(1+3\lambda)^2-\lambda^2}.\label{rho} 
\end{equation}

In this paper, we assume that the universe consists of non-relativistic pressureless matter, characterized by a pressure-density relation with $\gamma=1$. This assumption simplifies the analysis and allows us to focus on the effects of viscosity in a matter-dominated universe. From Eqs. (\ref{2e}), (\ref{F1}) and (\ref{F2}), we have
\begin{equation}
\overset{.}{H}+\frac{3\left( 2\lambda +1\right) }{2\left( 3\lambda +1\right) 
}H^{2}-\frac{3\zeta \left( 2\lambda +1\right) \left( 4\lambda +1\right) }{%
2\left( 3\lambda +1\right) }H=0.
\end{equation}

Now, by replacing $\frac{1}{H} \frac{d}{dt}$ with $\frac{d}{d\ln(a)}$, the above equation becomes
\begin{equation}
    \frac{dH}{d\ln \left( a\right) }+\frac{3\left( 2\lambda +1\right) }{2\left(
3\lambda +1\right) }H-\frac{3\zeta \left( 2\lambda +1\right) \left( 4\lambda
+1\right) }{2\left( 3\lambda +1\right) }=0.
\label{ed}
\end{equation}

By using the relation $a(t) = 1/(1 + z)$ and integrating Eq. (\ref{ed}), we obtain the following solution for the Hubble parameter in terms of redshift $z$,
\begin{equation}
H\left( z\right) =H_{0}\left( 1+z\right) ^{\frac{6\lambda +3}{6\lambda +2}%
}+\zeta \left( 4\lambda +1\right) \left[ 1-\left( 1+z\right) ^{\frac{%
6\lambda +3}{6\lambda +2}}\right].
\label{Hz}
\end{equation}

Setting $z = 0$ in Eq. (\ref{Hz}), we find that $H(0) = H_0$, where $H_0$ represents the present value of the Hubble parameter. Specifically, for the case where $\lambda = 0$ and $\zeta = 0$, this solution describes a universe dominated by non-relativistic matter, where the expansion rate $H(z)$ at redshift $z$ is proportional to $(1 + z)^{3/2}$, a characteristic behavior of matter-dominated eras in cosmology. In addition, the dynamics and fundamental cosmological characteristics of the model specified in Eq. (\ref{Hz}) are solely governed by the model parameters ($\lambda$, $\zeta$). In the following section, we constrain these parameters ($H_0$, $\lambda$, $\zeta$) using up-to-date observational datasets to explore the evolution of cosmological parameters.

\section{Observational constraints}
\label{sec5}

To investigate the observational implications of our viscous $f(R,T)$ cosmological model, we employ the latest cosmic Hubble and SNe observations. These observational datasets provide crucial information about the expansion history of the universe and the behavior of DE. By comparing the theoretical predictions of our viscous $f(R,T)$ cosmological model with these observations, we aim to constrain the model parameters and assess its compatibility with observational data. We use 31 data points from the Hubble $H(z)$ datasets and 1701 data points from the Pantheon+ supernova samples. We employ the Bayesian analysis, likelihood function, and Markov Chain Monte Carlo (MCMC) method provided by the emcee Python library \cite{Mackey/2013}.

\subsection{$H(z)$ datasets}

The Hubble parameter $H(z)$ describes the rate of expansion of the universe at redshift $z$. It can be expressed as $H(z) = -\frac{dz}{dt(1 + z)}$. This formula relates the change in redshift $z$ with respect to cosmic time $t$ and accounts for the cosmic expansion factor $(1+z)$. Since $dz$ is obtained from a spectroscopic survey, we can calculate the model-independent value of the Hubble parameter by measuring the quantity $dt$. In our analysis, we include a set of 31 data points obtained from the differential age approach to prevent additional correlation with BAO data \cite{Sharov/2018}. The best-fit values of the model parameters $H_0$, $\lambda$, and $\zeta$ are determined using the chi-square function $\chi^{2}$, calculated as
\begin{equation}
\Tilde{\chi}^{2}_{H(z)} = \sum_{i=1}^{31} \frac{\left[H_{th}(\theta_{s}, z_{i})-
H_{obs}(z_{i})\right]^2}{\sigma^2(z_{i})}.
\end{equation}

Here, $H_{th}(\theta_{s},z_{i})$ denotes the theoretical prediction of $H(z)$ at redshift $z_{i}$, and $H_{obs}(z_{i})$ epresents the corresponding observed values. The term $\sigma^{2}(z_{i})$ represents the standard error associated with the measured values of $H_{obs}(z_{i})$ and $\theta_{s}=(H_0,\lambda,\zeta)$ defines the parameter space of our viscous $f(R,T)$ cosmological model.

\subsection{Pantheon+ datsets}

Recent observational discoveries concerning SNe Ia have corroborated the existence of the accelerated expansion phase in our universe. In the past two decades, there has been a significant rise in the volume of observations of SNe Ia samples. The Pantheon dataset comprises 1048 SNe Ia samples, covering a redshift range of $0.01 < z < 2.3$. The dataset was published in 2018 \cite{Scolnic/2018}. This collection of observations includes data from various low-redshift surveys, as well as surveys conducted by the Hubble Space Telescope, the Pan-STARRS1 Medium and Deep Surveys, the Supernova Legacy Survey, and the Sloan Digital Sky Survey. Recently published \cite{Scolnic/2022,Brout/2022}, the Pantheon+ sample comprises 1701 light curves of 1550 Type Ia supernovae within the redshift range of $[0.001, 2.26]$. The luminosity distance is assumed to be \cite{Planck/2020},
\begin{equation}
    D_L(z)=\frac{c(1+z)}{H_0}S_K\bigg(H_0\int^z_0\frac{d\overline{z}}{H(\overline{z})}\bigg),
\end{equation}
where
\begin{equation}
S_K(x)=    \begin{cases}
      \sinh(x\sqrt{\Omega_K})/\Omega_K,\quad ~ \Omega_K >0\\
      x,\quad\quad\quad\quad\quad\quad\quad \quad ~~\Omega_K=0\\
      \sin (x \sqrt{|\Omega_K|})/|\Omega_K|,~~ \Omega_K<0
    \end{cases}\,.
\end{equation}
where $c$ represents the speed of light. In the case of a spatially flat universe, we have
\begin{equation}
D_L(z)=c(1+z)\int^z_0\frac{d\overline{z}}{H(\overline{z})}.
\end{equation}

Theoretically, the distance modulus can be formulated as
\begin{equation}
\mu^{th}=5\log_{10}D_L(z)+\mu_0,\quad \mu_0 = 5 \log_{10} \frac{1}{H_0 Mpc}+25.
\end{equation}

The chi-square function for the Pantheon+ samples is expressed as
\begin{equation}
\chi^2_{SNe}(\theta_s)=\sum_{i,j=1}^{N_{SNe}}\Delta\mu_{i}\left(C^{-1}_{SNe}\right)_{ij}\Delta\mu_{j}, \quad  \Delta\mu_i=\mu^{th}(\theta_s)-\mu_i^{obs},
\end{equation}
where $\mu^{th}$ represents the expected value of the distance modulus based on our viscous $f(R,T)$ cosmological model, while $\mu_i^{obs}$ represents its observed value. 

The empirical relation employed to calculate the distance modulus of SNe Ia from the observation of their light curves is expressed as $\mu= m_{B}-M_{B}+\alpha x_{1}-\beta c+ \Delta_{M}+\Delta_{B}$ \cite{Tripp/1998}. Here, $X1$ and $C$ represent the stretch and color correction parameters, respectively \cite{Scolnic/2022,Mukherjee/2021}. $m_B$ represents the observed apparent magnitude, and $M_B$ is the absolute magnitude in the B-band for SNe Ia. The parameters $\alpha$ and $\beta$ are two nuisance parameters that describe the luminosity stretch and luminosity color relations, respectively. In addition, the distance correction factor is denoted as $\Delta_{M}$, and $\Delta_{B}$ is a distance correction based on predicted biases from simulations. Using the BEAMS with Bias Correction (BBC) approach \cite{Kessler/2017,Fotios/2021}, we define the observed distance modulus as the difference between the apparent magnitude $m_{B}$ and the absolute magnitude $M_{B}$, denoted as $\mu = m_{B} - M_{B}$. We will not marginalize over the nuisance parameters $\alpha$ and $\beta$ but instead marginalize over the Pantheon+ data for $M_{B}$. Therefore, we do not consider the values of $\alpha$ and $\beta$ in the current investigation of the viscous $f(R,T)$ cosmological model.

Furthermore, the distance modulus's total uncertainty matrix is represented as
\begin{equation}
C_{SNe} = D_{stat}+C_{sys}.
\end{equation}

Also, we consider that $D_{ii,stat}=\sigma^2_{\mu(z_i)}$ represents the diagonal matrix of statistical uncertainties. Following Scolnic et al. \cite{Scolnic/2018}, the BBC approach is employed to determine systematic uncertainty,
\begin{equation}
C_{ij,sys} = \sum^K_{k=1}\bigg(\frac{\partial \mu^{obs}_i}{\partial S_k}\bigg)
\bigg(\frac{\partial \mu^{obs}_j}{\partial S_k}\bigg)\sigma^2_{S_k}.
\end{equation}

Here, $S_k$ represents the magnitude of the systematic error, $\sigma_{S_k}$ is its standard deviation uncertainty, and the indices ${i,j}$ refer to the redshift bins of the distance modulus.

Now, to derive combined constraints for the parameters $H_0$, $\lambda$, and $\zeta$ from the $H(z)$ and Pantheon+ samples, we use the total likelihood function. The corresponding likelihood and $\chi^2$ functions are defined as
\begin{equation}
\mathcal{L}_{joint} = \mathcal{L}_{H(z)} \times \mathcal{L}_{Pantheon+},    
\end{equation}
and
\begin{equation}
\chi^{2}_{joint} = \chi^{2}_{H(z)} + \chi^{2}_{Pantheon+},    
\end{equation}
where $\mathcal{L}_{H(z)}=\exp(-\chi^2_{H(z)}/2)$ and $\mathcal{L}_{Pantheon+}=\exp(-\chi^2_{Pantheon+}/2)$ are the likelihood functions for the $H(z)$ and $Pantheon+$ samples, respectively. We derived the constraints on the parameters of our viscous $f(R,T)$ cosmological model by minimizing the total chi-squared function using the combined $H(z)$ + Pantheon datasets. In our MCMC analysis, we impose the following priors: $H_0 \in [60,80], \lambda \in[-1,1], \text{and } \zeta \in [0,200]$. We use 100 walkers and 1000 steps to determine the fitting results. In Fig. \ref{CC+SNe}, we present the $1-\sigma$ and $2-\sigma$ likelihood contours for the model parameters $H_0$, $\lambda$, and $\zeta$. The best-fit values derived from the $1-\sigma$ and $2-\sigma$ contours depicted in Fig. \ref{CC+SNe} are: $H_0=67.8^{+1.1}_{-1.1}$, $\lambda=-0.1619^{+0.0045}_{-0.0044}$, and $\zeta=140.5^{+1.9}_{-2.0}$. Note that the free parameter $\lambda$ is dimensionless, while the bulk viscosity coefficient $\zeta$ has units of Pascal-seconds ($Pa.s$) in the SI system or cubic meters ($M^3$) in the Planck system. In addition, 
Fig. \ref{Hubble} displays the error bar plot comparing the viscous $f(R,T)$ cosmological model with the $\Lambda$CDM (standard cosmological model) using the cosmological parameters $\Omega_{m0} = 0.315$ and $H_0 = 67.4$ km/s/Mpc. The value of $H_0$ obtained is in agreement with measurements from the Planck mission \cite{Planck/2020} and recent studies \cite{Yang/2021,Valentino/2021B}. Further discussion on $H_0$ can be found in \cite{Valentino/2021}.

\begin{widetext}

\begin{figure}[h]
\centering
\includegraphics[scale=0.8]{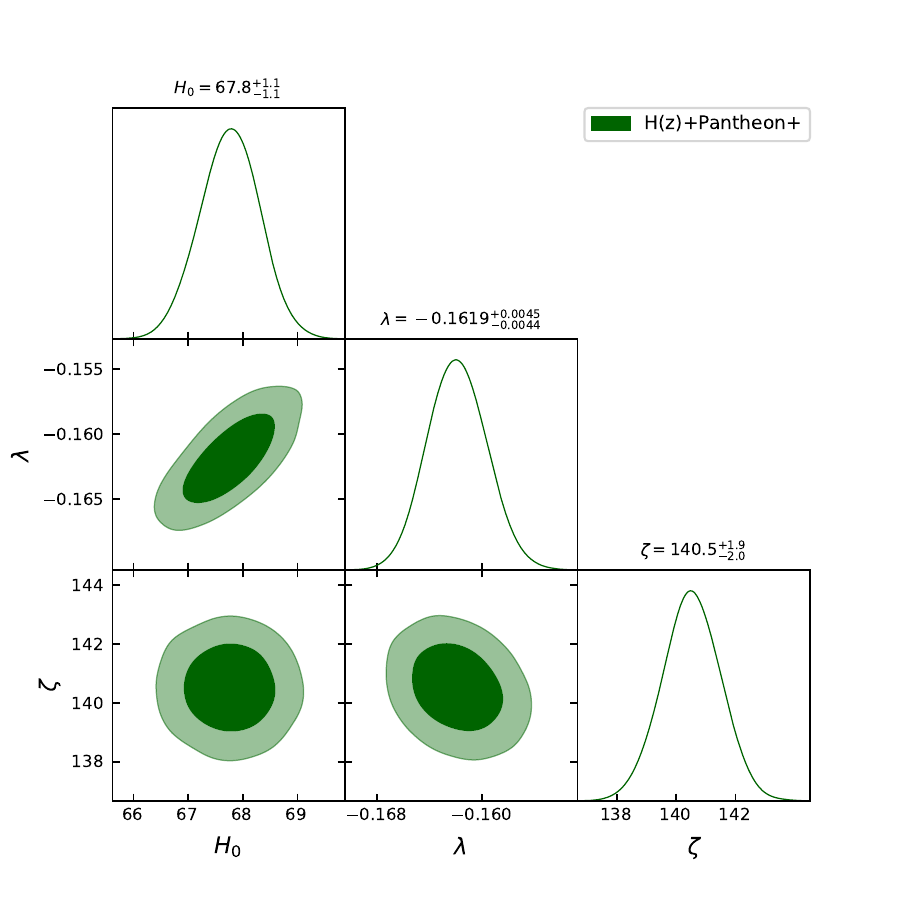}
\caption{The likelihood contours for the model parameters, shown as $1-\sigma$ and $2-\sigma$, are determined using the combined $H(z)$+Pantheon+ datasets.}
\label{CC+SNe}
\end{figure}

\begin{figure}[h]
\centering
\includegraphics[width=18cm,height=5.5cm]{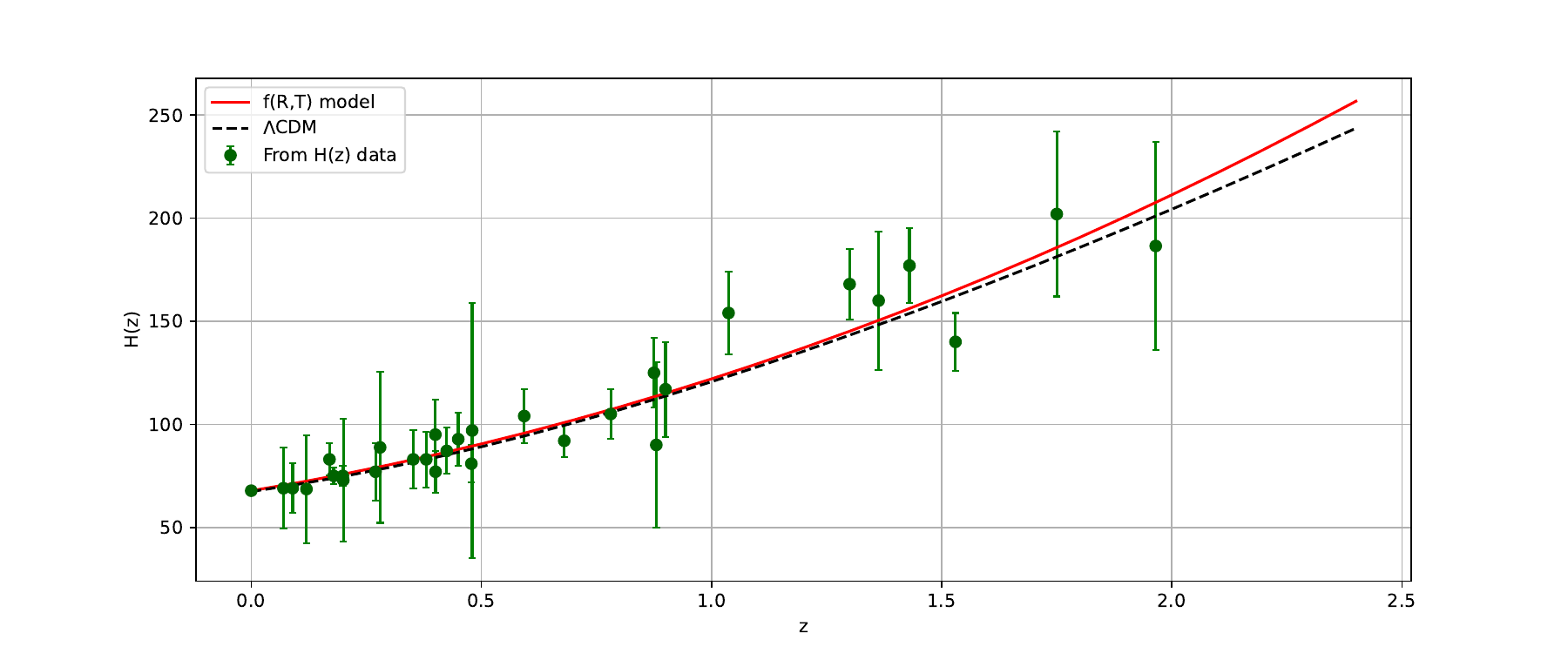}
\caption{The plot shows the error bars of $H(z)$ vs. $z$ for our viscous $f(R,T)$ cosmological model. The solid green line represents the curve for the viscous $f(R,T)$ cosmological model, while the black dotted line corresponds to the $\Lambda$CDM model. The green dots indicate the 31 points of the Hubble $H(z)$ datasets.}
\label{Hubble}
\end{figure}	

\end{widetext}

\section{Cosmological applications}
\label{sec6}

In contemporary cosmological literature, there is a prevalent trend to focus on mathematical solutions within modified gravitational theories, which may lack physical grounding \cite{Shamir/2015}. Our study critically examines this trend, ensuring our approach is firmly rooted in physical principles. By investigating the implications of modified gravity theories and viscosity on cosmological parameters such as energy density, pressure component with viscosity, and effective EoS parameter, we aim to offer a scientifically justified perspective on the universe's dynamics, contributing to a deeper understanding of cosmic evolution. These analyses are based on the best-fit values of the model parameters $H_0$, $\lambda$, and $\zeta$, which were constrained using the combined $H(z)$+Pantheon+ datasets, and then explored for different values of $\zeta$.

\begin{figure}[h]
   \begin{minipage}{0.48\textwidth}
     \centering
     \includegraphics[width=\linewidth]{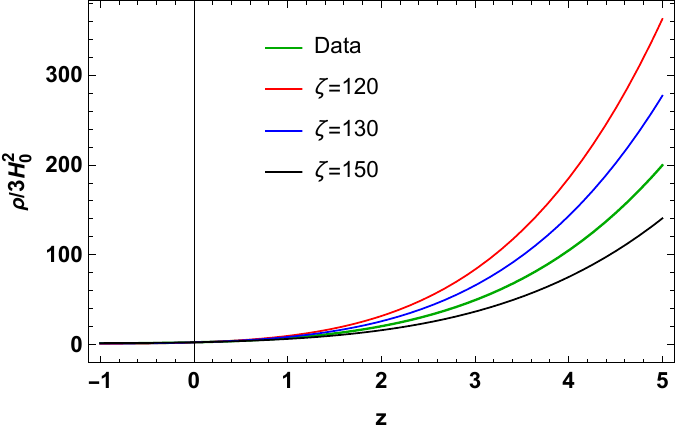}
     \caption{The energy density behavior for the specified model, based on the parameters constrained by the $H(z)$+Pantheon+ datasets and different values of $\zeta$.}\label{F_rho}
   \end{minipage}\hfill
   \begin{minipage}{0.48\textwidth}
     \centering
     \includegraphics[width=\linewidth]{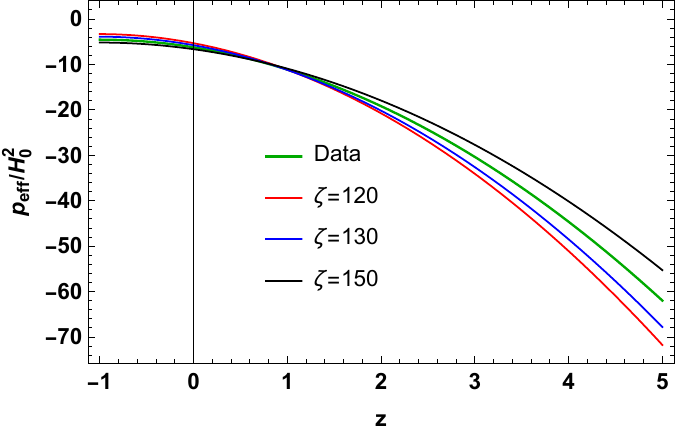}
     \caption{The effective pressure behavior for the specified model, based on the parameters constrained by the $H(z)$+Pantheon+ datasets and different values of $\zeta$.}\label{F_p}
   \end{minipage}
\end{figure}

First, we consider the energy density, which describes the total amount of matter present in the universe at different redshifts. From Fig. \ref{F_rho}, it is evident that the energy density exhibits a positive behavior for the constrained values of the model parameters. This behavior aligns with expectations, as the energy density is expected to decrease as the universe expands. It starts with significantly positive values and gradually approaches zero in the future, specifically at $z=-1$. This behavior is consistent with the standard cosmological model and provides further validation for our viscous $f(R,T)$ cosmological model.

Next, we analyze the pressure component that includes viscosity. Viscosity affects the behavior of matter in the universe, influencing its expansion and evolution. Fig. \ref{F_p} illustrates that the bulk viscous cosmic fluid demonstrates negative pressure across all redshift values. This property is significant because it suggests that bulk viscosity could be a viable candidate to drive the cosmic acceleration observed in the universe. The negative pressure exerted by the bulk viscous fluid contributes to the repulsive gravitational effect that leads to the expansion of the universe at an accelerating rate. This behavior is consistent with the characteristics expected from a DE component, further highlighting the potential of bulk viscosity as a mechanism for explaining cosmic acceleration in our viscous $f(R,T)$ cosmological model.

Furthermore, we examine the effective EoS parameter, which characterizes the relationship between effective pressure and energy density i.e. $\omega_{eff}=\frac{\bar{p}}{\rho}$. The effective EoS parameter for our viscous $f(R,T)$ cosmological model is given by
\begin{equation}
    \omega_{eff}=-\frac{3\zeta H}{\rho}=\frac{\zeta(3 \lambda +1) }{ \left[\zeta(4 \lambda +1) -H_0\right](1+z)^{\frac{6 \lambda +3}{6 \lambda +2}}-\zeta  (3 \lambda +1)}.
\end{equation}

\begin{figure}[h]
   \begin{minipage}{0.48\textwidth}
     \centering
     \includegraphics[width=\linewidth]{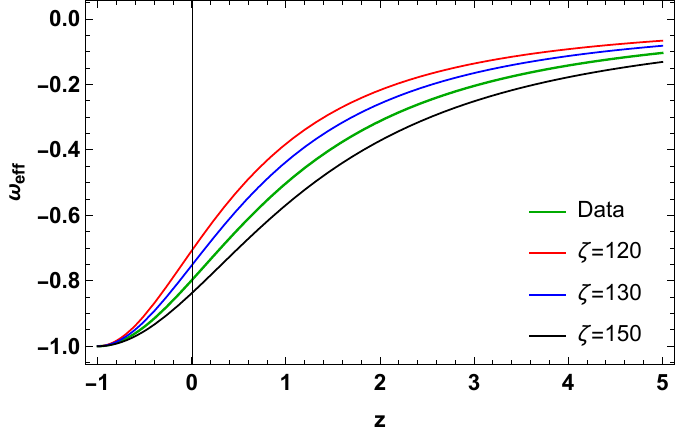}
     \caption{The effective EoS parameter behavior for the specified model, based on the parameters constrained by the $H(z)$+Pantheon+ datasets and different values of $\zeta$.}\label{F_EoS}
   \end{minipage}\hfill
   \begin{minipage}{0.48\textwidth}
     \centering
     \includegraphics[width=\linewidth]{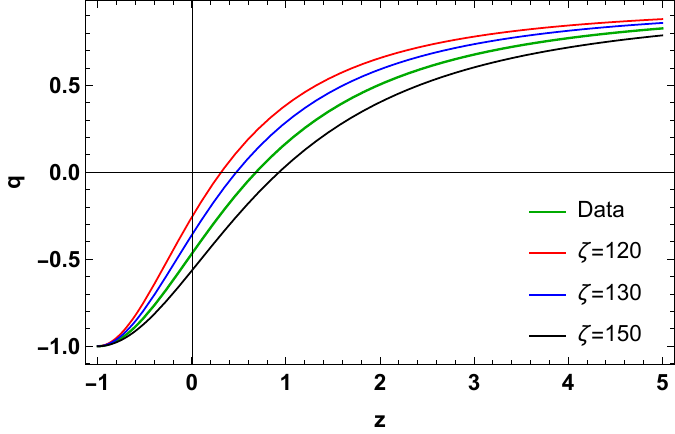}
     \caption{The deceleration parameter behavior for the specified model, based on the parameters constrained by the $H(z)$+Pantheon+ datasets and different values of $\zeta$.}\label{F_q}
   \end{minipage}
\end{figure}

The effective EoS parameter presented in Fig. \ref{F_EoS} indicates that the cosmic viscous fluid starts from a matter-dominated era ($\omega_{eff}=0$), crosses into the quintessence region ($\omega_{eff}>-1$), and finally approaches the $\Lambda$CDM model ($\omega_{eff}=-1$). Also, the current value of the effective EoS parameter from the combined $H(z)$+Pantheon+ datasets is $\omega_{0} \approx -0.80$ \cite{Almada/2019,Zhang/2010,Myrzakulova/2024,Koussour/2024,Koussour/2023}, indicating the universe's accelerating expansion phase and quintessence-like behavior.

Finally, the sign of the deceleration parameter, $q$, indicates the nature of the universe's expansion. When $q > 0$, the universe experiences decelerating expansion. If $q = 0$, the expansion rate is constant, and if $-1 < q < 0$, the expansion is accelerating. For $q = -1$, the universe exhibits exponential expansion or de Sitter expansion, while for $q < -1$, the expansion is super-exponential. The deceleration parameter for our viscous $f(R,T)$ cosmological model is given by
\begin{equation}
\label{q}
q=-1+\frac{d}{dt}\left( 
\frac{1}{H}\right)=-1+\frac{(6 \lambda +3) \left[H_0-\zeta  (4 \lambda +1)\right](1+z)^{\frac{1}{6 \lambda +2}+1}}{(6 \lambda +2) \left[\zeta(4 \lambda +1) + \left[H_0-\zeta  (4 \lambda +1)\right] (1+z)^{\frac{6 \lambda +3}{6 \lambda +2}}\right]}.
\end{equation}%

Fig. \ref{F_q} illustrates the transition of the deceleration parameter from a decelerated phase ($q > 0$) to an accelerated phase ($q < 0$) of the universe's expansion, ultimately approaching exponential expansion ($q = -1$) for the constrained values of the model parameters. The transition redshift from the combined $H(z)$+Pantheon+ datasets is approximately $z_t \approx 0.68$ \cite{Jesus/2020,Garza/2019}. The current value of the deceleration parameter is $q_0 \approx -0.47$ \cite{Mamon/2017,Mamon/2018,Myrzakulov1,Myrzakulova,Myrzakulov2} for the same datasets. In the absence of viscosity ($\zeta=0$), the deceleration parameter remains constant, specifically $q=\frac{1}{6 \lambda +2}$ (as shown in Ref. \cite{THK-7}). This leads to either a phase of constant acceleration or constant deceleration, depending on the value of $\lambda$. To account for both phases of expansion---acceleration and deceleration---we must include the effects of viscosity.

\section{Statefinder analysis}
\label{sec7}

The cosmological constant $\Lambda$ is plagued by two significant issues: the cosmological constant problem and the cosmic coincidence problem. To overcome these challenges, dynamic models of DE have been proposed in the literature, as we discussed in the Introduction. To distinguish between these varying DE models, a suitable tool was necessary. To address this issue, Sahni et al. \cite{Sahni/2003} introduced a new pair of geometrical parameters called statefinder parameters ($r, s$). The statefinder parameters are defined by the expressions:
\begin{equation}
r=\frac{\overset{...}{a}}{aH^{3}},\text{ \ \ }s=\frac{r-1}{3\left( q-\frac{1%
}{2}\right) }.  
\end{equation}

\begin{figure}[h]
   \begin{minipage}{0.48\textwidth}
     \centering
     \includegraphics[width=\linewidth]{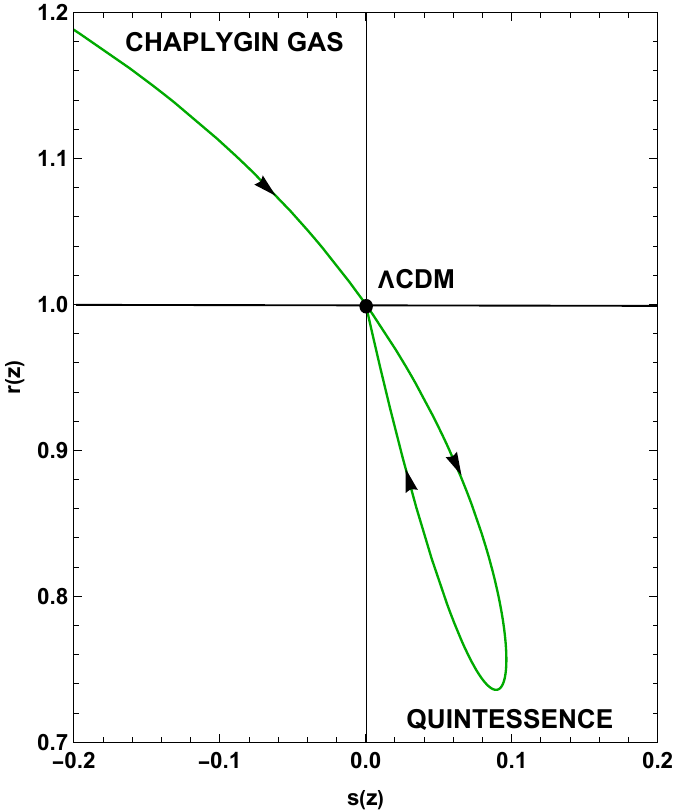}
     \caption{The $r-s$ plane behavior for the specified model, based on the parameters constrained by the $H(z)$+Pantheon+ datasets.}\label{F_rs}
   \end{minipage}\hfill
   \begin{minipage}{0.48\textwidth}
     \centering
     \includegraphics[width=\linewidth]{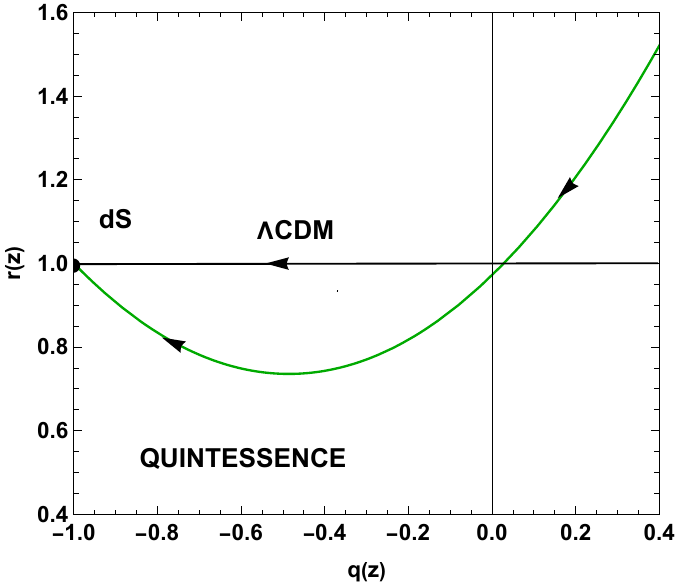}
     \caption{The $r-q$ plane behavior for the specified model, based on the parameters constrained by the $H(z)$+Pantheon+ datasets.}\label{F_rq}
   \end{minipage}
\end{figure}

For instance, the $\Lambda$CDM model, which features a cosmological constant, corresponds to the statefinder pair $(r, s) = (1, 0)$. On the other hand, models like the Chaplygin gas model, where the EoS transitions from a stiff fluid to a cosmological constant, are represented by $(r, s)$ values where $r > 1$ and $s < 0$. Quintessence models, which involve a scalar field driving the accelerated expansion of the universe, are characterized by $(r, s)$ values where $r < 1$ and $s > 0$. In Figs. \ref{F_rs} and \ref{F_rq}, we depict the $s-r$ and $q-r$ plots for our viscous $f(R,T)$ cosmological model using the parameter values constrained by the combined $H(z)$+Pantheon+ datasets. Figs. \ref{F_rs} and \ref{F_rq} illustrate that our viscous $f(R,T)$ cosmological model resides in the quintessence region. In addition, the evolutionary paths of our model deviate from the $\Lambda$CDM point. The current values of the statefinder parameters are $r_0 =0.74$ and $s_0 = 0.09$.

\section{Discussions and conclusions}
\label{sec8}

Cosmology has captivated the scientific community due to its focus on understanding the fundamental nature of the universe. DE, believed to be responsible for the accelerated expansion of the cosmos, challenges our understanding of fundamental physics. Dark matter (DM), which interacts gravitationally but not electromagnetically, comprises about 25\% of the universe's matter content and continues to elude direct detection, adding to the intrigue of the field \cite{cdmsii_collaboration/2010,akerib/2014,essig/2012}. Modified theories of gravity have also been employed to explain the dark sector of the universe \cite{bohmer/2008,mannheim/2012}. In such theories, the behavior attributed to DE and DM arises as a consequence of modifications to the gravitational laws, often involving changes to general relativity. These modified gravity theories offer a different perspective on the nature of the universe's acceleration and the gravitational interactions responsible for cosmic structures, potentially eliminating the need for exotic DE or DM components. Recently, efforts have been made to comprehend the results of the changes in spacetime around the exotic objects such as black holes, arising from their gravitational interaction with DM through numerical modeling. Thus, the accretion disk and the Quasi-Periodic oscillations are numerically computed to develop alternative solutions to observational results \cite{Donmez1,Donmez2,Donmez3,Paul:2023vys}.

In this paper, we investigated an extension of standard GR to describe DE using $f(R, T )$ modified theories of gravity. In these theories, the gravitational Lagrangian is expressed as an arbitrary function of the Ricci scalar $R$ and the trace of the stress-energy tensor $T$. The $f(R, T)$ gravity model, proposed by Harko et al. \cite{THK-7}, introduces a coupling between matter and geometry. This coupling results in a model where the gravitational field equations depend on a source term that represents the variation of the energy-momentum tensor with respect to the metric. This theory offers intriguing solutions that are particularly relevant in the fields of cosmology and astrophysics \cite{Myrzakulov/2012,Houndjo/2012,Barrientos/2014,Vacaru/2014,Jamil/2012,Sharif/2012,Sharif/2014,Silva,Vinutha,Bishi}. Our cosmological model based on $f(R,T)$ gravity considers a spatially homogeneous and isotropic flat metric, along with an energy-momentum tensor representing a viscous fluid. We adopt the simplest specific model, $f(R,T)=R+\lambda T$, where $\lambda$ is a constant. We then considered the effective EoS in Eq. (\ref{2e}), which corresponds to the Einstein case value with a proportionality constant $\zeta$, commonly employed in Einstein's theory \cite{Brevik/2012} and frequently referenced in the literature. From a hydrodynamic perspective, incorporating the viscosity coefficient into the cosmic matter content is a natural extension, as the ideal characteristics of a fluid are, fundamentally, an abstraction.

In Sec. \ref{sec4}, we derived the exact solution for our viscous $f(R,T)$ cosmological model and utilized 31 $H(z)$ data points from the differential age approach and 1701 points from the Pantheon+ samples. The best-fit values for the model's free parameters were obtained (see Fig. \ref{CC+SNe}), resulting in $H_0=67.8^{+1.1}_{-1.1}$, $\lambda=-0.1619^{+0.0045}_{-0.0044}$, and $\zeta=140.5^{+1.9}_{-2.0}$ for the combined $H(z)$+Pantheon+ datasets. Fig. \ref{Hubble} illustrates the comparison between our model's Hubble parameter and the cosmological data, contrasting it with the predictions of the $\Lambda$CDM model. The viscous $f(R,T)$ cosmological model exhibits a good agreement with observations, particularly at higher redshifts, where it demonstrates a superior fit compared to the $\Lambda$CDM model.

Furthermore, we analyzed the evolution of energy density, the pressure component considering viscosity, the effective EoS parameter, and the deceleration parameter as functions of redshift. These analyses are depicted in Figs. \ref{F_rho}-\ref{F_q}, using the model parameters constrained by the combined $H(z)$+Pantheon+ datasets and different values of $\zeta$. 
Fig. \ref{F_rho} demonstrates the expected positive behavior of the energy density. In addition, Fig. \ref{F_p} shows that the viscous fluid exhibits negative pressure at all redshift values, making bulk viscosity a viable candidate for driving cosmic acceleration. This behavior is further reflected in the effective EoS parameter behavior in Fig. \ref{F_EoS}, which indicates the universe's accelerating expansion phase and quintessence-like behavior. Further, the deceleration parameter presented in Fig. \ref{F_q} indicates the transition of the universe's expansion from a decelerated phase to an accelerated phase, ultimately approaching exponential expansion ($q = -1$). This transition is a characteristic feature of many cosmological models and is consistent with the observed acceleration of the universe's expansion.

Finally, Figs. \ref{F_rq} and \ref{F_rs} illustrate that the evolutionary paths of our viscous $f(R,T)$ cosmological model deviate from the fixed point of $\Lambda$CDM, where $r = 1$ and $s = 0$. In the current epoch, these paths lie in the quintessence region where $r < 1$ and $s > 0$. This indicates that our viscous $f(R,T)$ cosmological model provides a viable alternative to explain the dynamics of the universe, particularly without the need to invoke the cosmological constant. It's important to note that $f(R,T)$ gravity has demonstrated potential, further investigation is necessary to establish it as a viable gravitational formalism. In a study by several authors, it was demonstrated that $f(R,T)$ gravity can be compatible with constraints from the Solar System \cite{Shabani/2014,Bertini/2023}. Alves et al. \cite{Alves/2016} investigated the generation of gravitational waves within the frameworks of $f(R,T)$ and $f(R,T^\phi)$ theories of gravity. They explored the impact of these theories on the propagation of gravitational waves and discussed the potential observational implications of their findings.

\section*{Data availability} 
This article does not include any new associated data.

\section*{Acknowledgments} 
This research was funded by the Science Committee of the Ministry of Science and Higher Education of the Republic of Kazakhstan (Grant No. AP14972745).

\end{document}